\pdfoutput=1
\documentclass[aps,prl,floatfix,twocolumn,reprint]{revtex4-1}
\usepackage{physics}
\usepackage{amsfonts}
\usepackage{mathrsfs}
\usepackage{amsmath}
\usepackage{color}
\usepackage{graphicx}
\usepackage{bm}
\usepackage{amssymb}
\usepackage{xspace}
\usepackage{epstopdf}
\usepackage{dcolumn}
\usepackage{longtable}
\usepackage[colorlinks=true, letterpaper=true, pdfstartview=FitV, linkcolor=blue, citecolor=blue, urlcolor=blue]{hyperref}
\makeatletter

\newcommand{\Rmnum}[1]{\expandafter\@slowromancap\romannumeral #1@}
\makeatother
\DeclareMathAlphabet\mathbfcal{OMS}{cmsy}{b}{n}

\begin{document}
\title{Quantum Hall Charge Kondo Criticality}
\author{Zhi-qiang Bao}
\affiliation{Department of Physics, University of Texas at Dallas, Richardson, Texas 75080, USA}
\author{Fan Zhang}\email{zhang@utdallas.edu}
\affiliation{Department of Physics, University of Texas at Dallas, Richardson, Texas 75080, USA}
\date{\today}
\begin{abstract}
The long-thought charge Kondo effects have recently been experimentally realized
in the quantum Hall regime. This experiment, 
supported by numerics, 
exemplifies the realization of two-channel Kondo state, a non-Fermi Liquid,
and its crossover to the one-channel counterpart, a Fermi liquid.
Scaling up such a platform, we find a hierarchy of non-Fermi Liquids
and their tunable crossovers based on a renormalization group analysis.
Utilizing results from a conformal field theory,
we further examine the universal conductances of this strongly correlated system and their finite temperature scaling,
which elucidate the sharp distinctions between charge and spin Kondo physics.
\end{abstract}
\maketitle

\indent\textcolor{blue}{\em Introduction.}---The Kondo effect is one of the most fundamental problems in condensed matter physics~\cite{PW}.
Not only does it reveal how delocalized electrons interact with a local degeneracy,
but it also exemplifies scaling and universality in strongly correlated systems that are driven by weak coupling instabilities.
The spin Kondo effects and their universal scaling with $T/T_K$,  
where $T_K$ is an emergent Kondo temperature,
have been observed in metals~\cite{S0}, quantum dots~\cite{S1,S2,S3}, 
carbon nanotubes~\cite{S4,S5,S6}, and individual molecules~\cite{S7,S8,S9}.~Fascinatingly,~orbital~\cite{OK,OK1,OK2}, 
charge~\cite{Matveev1,Garate,LF}, and even Majorana~\cite{LF,Beri,Egger,Bao} degrees of freedom 
can also play the roles of spin in lending local degeneracies and producing Kondo effects.

Many strongly interacting systems can be described by Landau's Fermi-liquid (FL) theory.
A prime example is the one-channel Kondo (1CK) effect, 
in which a spin impurity is antiferromagnetically coupled to one reservoir of conduction electrons. 
The 1CK ground state, although complex, 
is a FL in which the spin impurity is exactly screened and the low-lying excitations are still quasiparticles~\cite{S0}.
However, the most intriguing problems in condensed matter physics arise when the FL theory breaks down,
such as Tomonaga-Luttinger liquid, fractional quantum Hall states, high-temperature superconductors, 
and multi-channel Kondo models~\cite{Bao,Beri,Egger,Blandin,AZ,Affleck1,DGG1,DGG2,Borda,Matveev2,Pierre,Sela,Fiete}.

Consider the spin two-channel Kondo (2CK) effect as an example. 
Two independent reservoirs of conduction electrons compete to screen a spin impurity and thus overscreen it, 
leaving an emergent spin impurity to be overscreened for the next stage.
Like a domino effect, all electrons are strongly frustrated, 
leading to non-Fermi-liquid (NFL) quantum criticality with the emergence of a decoupled Majorana~\cite{Kivelson}. 
Yet, the delicate 2CK effect is unlikely to occur, as any channel asymmetry relieves the frustration 
and drives the system toward the 1CK effect in the more strongly coupled channel~\cite{Blandin}. 

Not until recently were the elusive 2CK NFL and its crossover to the 1CK FL 
observed in two ingenious experiments~\cite{DGG2,Pierre}.
While one experiment~\cite{DGG2} realized the spin version in a double quantum-dot device, 
the other experiment~\cite{Pierre}, in the spin-polarized integer quantum Hall regime, 
utilized two degenerate macroscopic charge states of a metallic island to play the role of pseudospin.
One may wonder whether the more exotic higher-level Kondo NFL's and their NFL-NFL crossovers 
can be realized in similar experiments, and if so, 
what are their measurable fingerprints of such quantum criticality. 

\begin{figure}[b!]
\includegraphics[width=0.92\columnwidth]{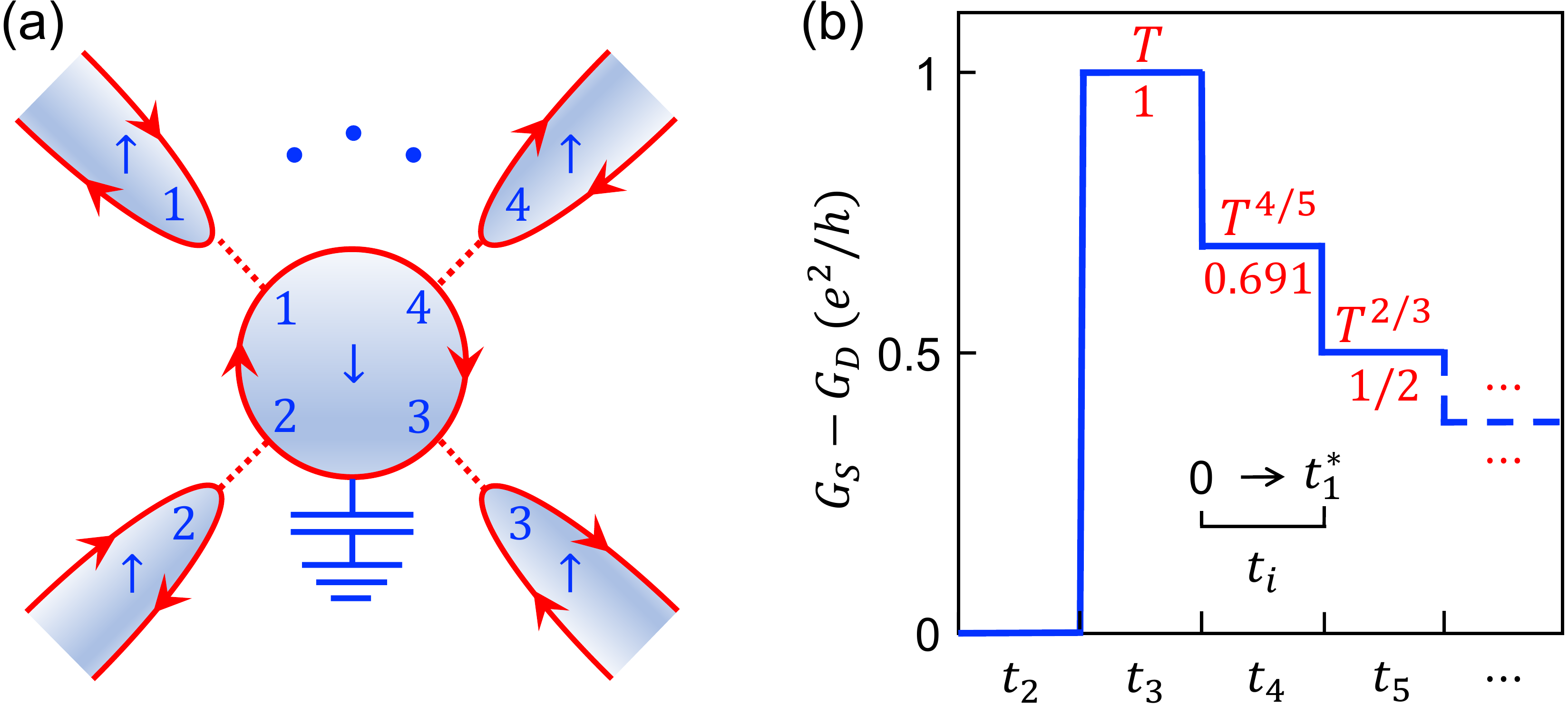}
\caption{(a) Schematic of  our quantum-Hall charge Kondo device, as described in the text. 
(b) Crossover between different Kondo states, as marked by their conductances and temperature scaling. 
A $t_{i}$-segment realizes an $(i\!-\!1)$-channel Kondo state; 
$t_{i}$ is tuned from $0$ to $t_1^\ast$ while $t_{j<i}=t_1^\ast$ and $t_{j>i}=0$.
At $t_{i}=t_1^\ast$, the crossover is between the 1CK FL and 2CK NFL for $i=2$,
but between two different NFL's for $i>2$.}
\label{fig1}
\end{figure}

\begin{figure}[t!]
\includegraphics[width=0.96\columnwidth]{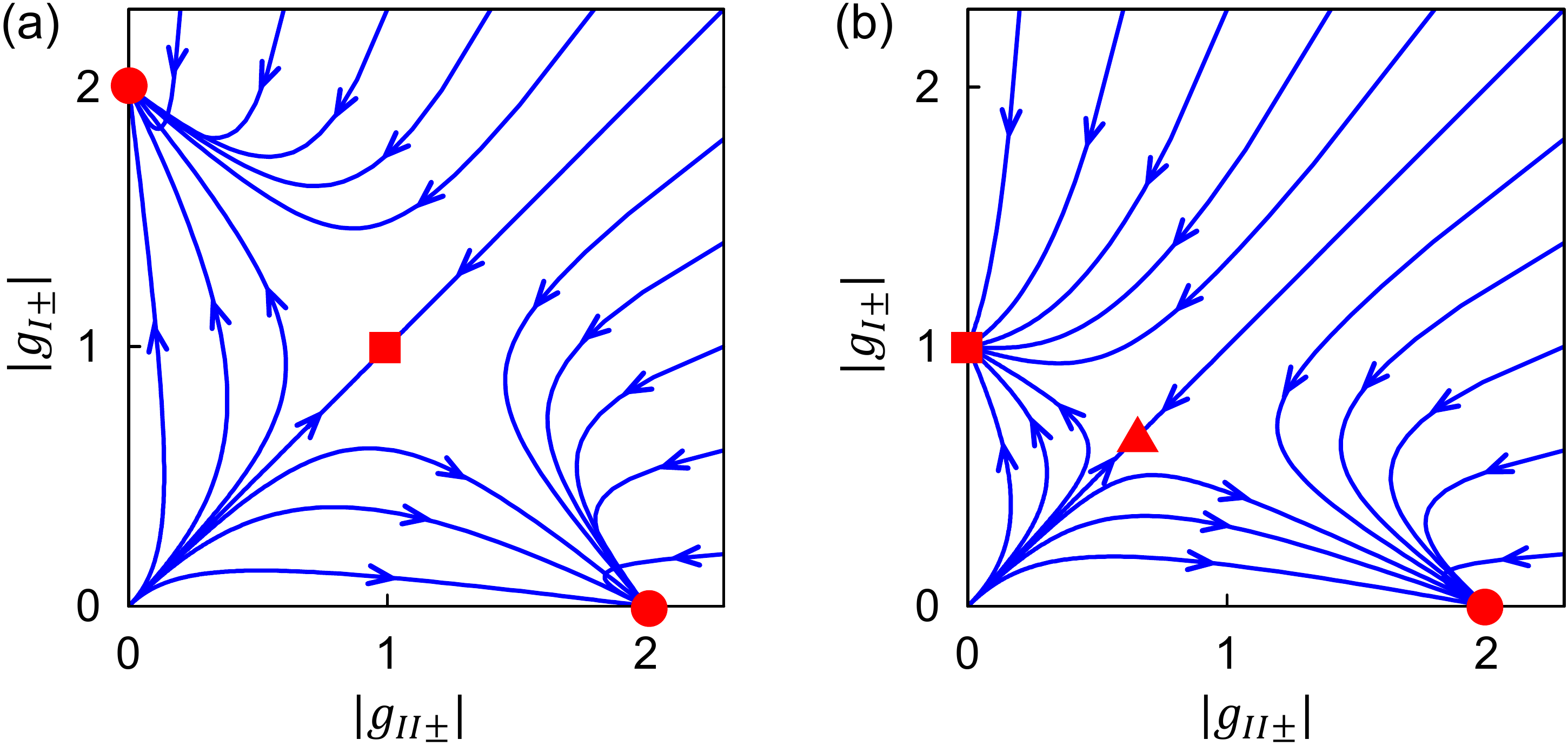}
\caption{RG flows projected onto the $g_{\Rmnum{1}\pm}$-$g_{\Rmnum{2}\pm}$ plane.
(a) The case for $N=2$ and $M=1$; (b) the case for $N=3$ and $M=2$.
The red dot, square, and triangle denote the 1CK (FL), 2CK (NFL), and 3CK (NFL) fixed points, respectively.}
\label{fig2}
\end{figure}

In this work, by exploiting the tunability and scalability of the quantum-Hall charge Kondo experiment~\cite{Pierre,Pierre2},
we demonstrate the realization of a hierarchy of Kondo NFL's and their tunable NFL-NFL crossovers (Fig.~\ref{fig1})
based on our systematic renormalization group (RG) analysis (Fig.~\ref{fig2}).
Moreover, although a boundary conformal field theory (BCFT) was originally applied 
to solving the spin multi-channel Kondo problems~\cite{Affleck1},
the universality of Kondo criticality enables us to use the correlation functions~\cite{Affleck2} obtained by the same BCFT
to compute analytically the zero-temperature conductances (Table~\ref{table1}) 
and their finite-temperature scaling (Table~\ref{table2})
in our proposed quantum-Hall charge Kondo device.
Importantly, the obtained transport anomalies immediately distinguish the charge and spin Kondo physics 
from a fundamental perspective. When applied to special cases, our results not only elucidate 
several key observations in the prototype experiment~\cite{Pierre} and its state-of-the-art simulation~\cite{Sela}
but also recover the remarkable results obtained indirectly from the quantum brownian motion~\cite{Yi1,Yi2}.
Our study provides new avenues for exploring Kondo criticality, 
quantum Hall effects, mesoscopic physics, and their tantalizing combinations.

\indent\textcolor{blue}{\em Model and RG analysis.}---We start from a description of our proposed mesoscopic device in Fig.~\ref{fig1}(a),
a direct generalization of the ingenious one~\cite{Pierre} in which the 2CK NFL 
and its crossover to the 1CK FL were observed.
First, the system is in the $\nu=1$ quantum Hall regime;
the central island and surrounding leads are characterized by one spinless chiral edge channel each,
and the leads are independently connected to the island via single-channel quantum point contacts (QPC). 
Second, the island has a capacitance $C$ such that the charging energy $E_c=e^2/2C$ 
($\sim 290$~mK in ref.~\onlinecite{Pierre}) is larger than the thermal energy $k_{\rm B}T$; 
this prevents charge buildup in the island, reducing its accessible charge states
to the pseudospin-$1/2$ manifold and strongly correlating tunneling events.
Third, the level spacing in the island $\delta$ ($\sim 0.2$~$\mu$K in ref.~\onlinecite{Pierre}) 
is much smaller than  $k_{\rm B}T$; 
the charge pseudospin flips when an electron is tunneled into or out of the island.

In light of the above analysis, the effective Hamiltonian governing the mesoscopic physics in Fig.~\ref{fig1}(a) is
\begin{eqnarray}\label{H}
H&=&\sum_{i=1}^N\sum_{\sigma=\uparrow,\downarrow}iv_{F}\int_{-\infty}^{\infty} dx
\psi_{i\sigma}^{\dagger}(x)\nabla_x\psi_{i\sigma}(x)\nonumber\\
&+&\sum_{i=1}^N\frac{t_{i}}{2}\left(\psi_{i\uparrow}^{\dagger}(0)\psi_{i\downarrow}(0)S^{-}
+\psi_{i\downarrow}^{\dagger}(0)\psi_{i\uparrow}(0)S^{+}\right),
\end{eqnarray}
where $\sigma=\;\uparrow$~and~$\downarrow$ denote the spinless electrons 
in the leads and in the central island, respectively,
$v_F$ is the edge state velocity, $x=0$ denotes the location of QPC's,
and each QPC is described by an independent electron tunneling $t_i>0$. 
Markedly, the Kondo exchange is first order in $t_i$'s,
as pseudospin flips are caused by single tunnelings through the QPC's.
This is in sharp contrast to the spin Kondo case, 
in which first-order events are forbidden and spin flips are virtual second-order events.    
This difference gives rise to fundamental distinctions in their transport behaviors, as will be shown later. 

Now we use the perturbative RG to analyze the flows of Kondo couplings and the influence of channel asymmetry.
For the latter purpose, we assume
$t_{i}=t_{\Rmnum{1}}$ for $1\leq i\leq M$ and $t_{i}=t_{\Rmnum{2}}$ for $M<i\leq N$.
By computing the standard one- and two-loop Feynman diagrams~\cite{Anderson,Zawadowski,Lindgren1,Lindgren2,Kuramoto1,Kuramoto2},
we derive the following RG flow equations:
\begin{eqnarray}\label{flow}
\!\frac{dg_{\alpha\pm}}{dl}\!&=&\!\frac{g_{\alpha\pm}}{4}\!\left[4g_{\alpha z}\!-\!M\!\left(g_{\Rmnum{1}z}^{2}\!+\!g_{\Rmnum{1}\pm}^{2}\right)\!-\!(N\!\!-\!\!M)\!\left(g_{\Rmnum{2}z}^{2}\!+\!
g_{\Rmnum{2}\pm}^{2}\right)\right]\nonumber\\
\frac{dg_{\alpha z}}{dl}\!&=&\!g_{\alpha\pm}^{2}\!-\!\frac{g_{\alpha z}}{2}\left[Mg_{\Rmnum{1}\pm}^{2}\!+\!
(N\!-\!M)\,g_{\Rmnum{2}\pm}^{2}\right],
\end{eqnarray}
where $\alpha=\Rmnum{1}\;\mbox{and}\;\Rmnum{2}$, $g_{\alpha\pm}=t_{\alpha}\nu$,
and $\nu$ is the edge-state density of states.
Note that Eq.~(\ref{H}) has no $g_{iz}s_{iz}S_{z}$ terms,
yet they can develop upon renormalization. 
For the symmetric case in which $t_{\Rmnum{1}}=t_{\Rmnum{2}}$, all the coupling constants flow to $2/N$, 
i.e., the intermediate coupling fixed point of $N$-channel Kondo states. By contrast, for the asymmetric case, 
$(g_{\Rmnum{1}\pm},g_{\Rmnum{1}z};g_{\Rmnum{2}\pm},g_{\Rmnum{2}z})$ flow to $(2/M,2/M;0,0)$
if $t_{\Rmnum{1}}\!>\!t_{\Rmnum{2}}$ and to $(0,0;2/(N\!\!-\!\!M),2/(N\!\!-\!\!M))$ if $t_{\Rmnum{1}}\!<\!t_{\Rmnum{2}}$,
realizing the $M$- and $(N\!\!-\!\!M)$-channel Kondo states, respectively.
Our result is in harmony with the Nozi\`{e}res-Blandin one~\cite{Blandin} for the case of $N=2$ and $M=1$:
any channel asymmetry in the delicate 2CK effect relieves the NFL and 
drives the system toward the 1CK effect in the more strongly coupled channel.
To the best of our knowledge, for the first time we generalize
this celebrated result to the case with an arbitrary number of channels:
the Kondo fixed point is determined solely by the number of channels with the most strongest coupling
while any other channel with a weaker coupling decouples completely.
To illustrate this result, the projected RG flows in two representative cases are plotted in Fig.~\ref{fig2}.
Whereas Fig.~\ref{fig2}(a) features two 2CK NFL to 1CK FL crossovers for $N=2$,
Fig.~\ref{fig2}(b) exhibits the 3CK NFL to 1CK FL and to 2CK NFL crossovers for $N=3$.

While the 2CK NFL to 1CK FL crossover has already been observed in the recent charge Kondo experiment~\cite{Pierre},
here we propose a similar experiment in which a hierarchy of NFL-NFL crossovers, 
which are more exotic and yet to be realized, can be established.
Consider the experiment sketched in Fig.~\ref{fig1}(b). 
Since each QPC in Fig.~\ref{fig1}(a) can be controlled externally and independently, 
we tune $t_{i}$ from $0$ to $t_1^\ast$ one by one (from $i=2$) while keeping all $t_{j<i}=t_1^\ast$ and all $t_{j>i}=0$.
According to our RG results, we can conclude that the quantum-Hall charge Kondo device realizes an $(i\!-\!1)$-channel Kondo state 
during the course of tuning $t_i$ and crosses over to the $i$-channel Kondo state once $t_{i}$ reaches $t_1^\ast$.
Markedly, the proposed setup in Fig.~\ref{fig1} is indeed a fertile ground for realizing and tuning NFL-NFL crossovers; 
the crossover is always between two different NFL's for any case with $i>2$.
(The device in ref.~\onlinecite{Pierre} has three QPC's and can be readily used for examining the $i=3$ case.)
Below, we will compute the universal conductances mediated by these NFL's
and how their finite-temperature scaling laws deviate from the standard $T^2$ law of FL's.

\indent\textcolor{blue}{\em Zero-temperature conductances.}---Now we consider the multi-terminal conductances
across the NFL states that we have identified. The charge current in the lead $i$ reads
\begin{align}\label{Ji}
J_i(z)=ev_F\left[n_{i\uparrow}(z)-n_{i\uparrow}(\bar z)\right]=e\left[J_i^s(z)-J_i^s(\bar z)\right],
\end{align}
where $n_{i\sigma}=\psi^\dag_{i\sigma}\psi_{i\sigma}$ is the pseudospin density,
$J_i^s=v_F$ $(n_{i\uparrow}-n_{i\downarrow})/2$ is the pseudospin current,
$z=\tau+ix$ with $x>0$ is the $(1+1)$-dimensional coordinate.  
The second equality in Eq.~(\ref{Ji}) reflects that it is the Kondo coupling that flips the pseudospin.
From the Kubo formula, the linear conductance tensor can be written as
\begin{align}\label{Gij}
\!\!{G}_{ij}=\lim_{\omega\to0}\frac{-1}{\hbar\omega L}
\int_{-\infty}^{\infty}d\tau e^{i\omega\tau}\int_{0}^{L}dx'
\langle \mathcal{T}_{\tau}J_{i}(z)J_{j}(z')\rangle,
\end{align}
where $z=\tau+ix$ and $z'=0+ix'$.

The conductance tensor can be simplified by considering the symmetry and renormalization of the Kondo couplings.
Without loss of generality, we focus on $t_{\Rmnum{1}}>t_{\Rmnum{2}}$ case,
in which the first $M$ leads have the same and relatively stronger coupling strength.
First of all, the $N-M$ leads with weaker couplings decouple the central island at the NFL fixed points,
hence $\langle J_iJ_j\rangle\sim\langle J_i\rangle\langle J_j\rangle=0$
and $G_{ij}=0$ if $i\;\mbox{or}\;j> M$. 
Secondly, as the $M$ leads with stronger couplings have an $S_{M}$ symmetry,
$G_{ij}$ can be denoted by the same and different terminal conductances $G_S$ and $G_D$ as follows:
\begin{flalign}\label{GSD}
{G}_{ij}=\left\{
\begin{array}{cc}
                (G_S-G_D)\delta_{ij}+G_D &  (i\;\mbox{and}\;j\leq M) \vspace{0.05in}\\
                0 & (i\;\mbox{or}\;j> M) \\
\end{array}. \right.
\end{flalign}

\begin{table}[t!]
\centering
\caption{\label{table1}
{The comparison of zero-temperature conductances in units of $e^2/h$
in the $M$-channel quantum-dot spin Kondo and quantum-Hall charge Kondo transports.
In the former, the source-drain asymmetry prefactor~\cite{DGG2} is not included.}}
\newcommand\T{\rule{0pt}{3.1ex}}
\newcommand\B{\rule[-1.7ex]{0pt}{0pt}}
\centering
\begin{tabular}{ccccc}
\hline\hline
M & 2 & 3 & 4 & M \T\\[3pt]
\hline
Spin Kondo & $\;1\;$ & $\;0.618\;$ & $\;0.423\;$ & $\;1\!-\!{\cos\frac{2\pi}{M+2}}{\sec\frac{\pi}{M+2}}$~\cite{Affleck1} \T\\[6pt]
Charge Kondo & $\;1\;$ & $\;0.691\;$ & $\;0.5\;$ & $\;2\sin^{2}\frac{\pi}{M+2}$~\cite{Yi1}\\[3pt]
\hline\hline
\end{tabular}
\end{table}

Furthermore, the $S_{M}$ symmetry dictates
the conductance tensor $G$ to have $M-1$ eigenvalues as $G_S-G_D$ and one eigenvalue as $(G_S-G_D)+MG_D$.
Since the total current flowing into the central island must be zero,
$\sum_{i}G_{ij}=\sum_{j}G_{ij}=0$, it follows that
\begin{align}\label{G0}
(G_S-G_D)+MG_D=0.
\end{align}
As the nontrivial block of $G$ is real and symmetric,
there exists an orthogonal matrix $\mathcal{Q}_{M\times M}$ that diagonalizes it.
It follows that $\mathcal{J}_{i}=\sum_{j=1}^{M}\mathcal{Q}_{ij}J_{j}^{s}$,
and that the trivial eigenvalue concluded in Eq.~(\ref{G0}) must correspond to the total spin current,
i.e., $\mathcal{J}_M=\sum_{j=1}^{M}J_{j}^{s}/\sqrt{M}$.

Therefore, the conductance tensor is in fact characterized by a single constant $G_S-G_D$.
Such a hallmark of the quantum-Hall charge Kondo criticality is related to
the two-point correlation functions of $\mathcal{J}_i$ with $i<M$.
These correlators, though never used for the spin Kondo problems,
characterize the NFL fixed points and
can be computed using the BCFT~\cite{Affleck2} as follows:
\begin{align}\label{JJ}
\langle\mathcal{T}_{\tau}\mathcal{J}_{i}(z)\mathcal{J}_{j}(z')\rangle =\frac{\delta_{ij}C}{8\pi^{2}(z-z')^{2}}\,.
\end{align}
For the case of $xx'>0$, the correlator does not involve any scattering with the NFL at $x=0$,
and naturally $C=1$ in the free electron theory. By sharp contrast,
for the case of $xx'<0$, the correlator is strongly modified by scattering with the NFL at $x=0$,
and it turns out $C=({S^{1}_{1/2}/S^{1}_{0}})/({S^{0}_{1/2}/S^{0}_{0}})$ in the BCFT,
where $S_{\mu}^{\nu}\!=\!\sqrt{{2}/({M+2})}\sin\left[{\pi(2\mu+1)(2\nu+1)}/({M+2})\right]$~\cite{Kac}.
By carrying out the integral in Eq.~(\ref{Gij}) using Eq.~(\ref{JJ}) after the diagonalization, we obtain
\begin{align}\label{G1}
G_S-G_D=\frac{2e^{2}}{h}\sin^{2}\frac{\pi}{M+2}\,.
\end{align}

Our results~(\ref{G0})~and~(\ref{G1}) based on the $S_M$ symmetry and the BCFT
completely solve the conductance tensor~(\ref{Gij}).
Table~\ref{table1} summarizes such characteristic zero-temperature conductances 
across the NFL's in the quantum-Hall charge Kondo criticality.
A few comments are in order.
(i) The $N=M=2$ charge Kondo experiment~\cite{Pierre} has already confirmed that $G_S-G_D={e^{2}}/{h}$.
Our predicted $0.691 e^{2}/{h}$ in the $M=3$ case and $e^{2}/{2h}$ in the $M=4$ case can be verified by similar experiments.
(ii) The ${e^{2}}/{h}$ conductance has also been obtained in spin and Majorana 2CK effects~\cite{Borda,DGG1,DGG2,Bao}
(if the asymmetry prefactor~\cite{DGG2} is ignored).
We note that this is a coincidence rather than a similarity;
as shown below, their finite-temperature scaling behaviors differ greatly.
(iii) In the symmetric limit ($M=N$), Eq.~(\ref{G1}) coincides with the Yi-Kane conductance~\cite{Yi1,Yi2},
which was mapped to the mobility of a less familiar quantum brownian motion in a nonsymmorphic lattice. 
Our approach is more direct and naturally facilitates the study of temperature scaling of $G_{ij}$.

\begin{table}[t!]
\centering
\caption{\label{table2}
{The comparison of finite-temperature scaling behaviors of conductances
in the  $M$-channel quantum-dot spin Kondo and quantum-Hall charge Kondo transports.
Any deviation from the FL $T^2$-law implies an NFL behavior.}}
\newcommand\T{\rule{0pt}{3.1ex}}
\newcommand\B{\rule[-1.7ex]{0pt}{0pt}}
\centering
\begin{tabular}{ccccc}
\hline\hline
M & 2 & 3 & 4 & M \T\\[3pt]
\hline
Spin Kondo & $\;T^{1/2}$~\cite{Bao,Affleck1,Borda,DGG1,DGG2}$\;$ & $\;T^{2/5}\;$ & $\;T^{1/3}\;$ & $\;T^{2/(M+2)}$~\cite{Affleck1} \T\\[6pt]
Charge Kondo & $\;T$~\cite{Matveev2,Pierre,Sela} & $\;T^{4/5}\;$ & $\;T^{2/3}\;$ & $\;T^{4/(M+2)}$\\[3pt]
\hline\hline
\end{tabular}
\end{table}

\indent\textcolor{blue}{\em Finite-temperature scaling.}---The behaviors of a NFL at low temperature
are determined by the leading irrelevant operator $\mathbfcal{J}_{-1}\cdot{\bm\phi}$ that has a dimension 
$1+\Delta$ with $\Delta=2/(M+2)$~\cite{Affleck1}.
Thus, the low temperature dependence of the conductance can be derived using a perturbation theory. 
As $\mathbfcal{J}_{-1}\cdot{\bm\phi}$ and $\mathcal{J}_i$ $(i<M)$ are both Virasoro primary operators~\cite{Affleck8},
the first-order correction 
$\sim\left\langle\mathcal{T}_{\tau}\mathcal{J}_{i}(z_{1})\mathcal{J}_{j}(z_{2})\mathbfcal{J}_{-1}\cdot{\bm\phi}(z_{3})\right\rangle$ vanishes.
As a result, the leading correction must be the second-order one, i.e.,  
\begin{align}\label{GT}
&G_{ij}(T)-G_{ij}(0)\propto\int_{-\frac{\beta}{2}}^{\frac{\beta}{2}}d\tau\int_{0}^{L}dx'\int_{-\frac{\beta}{2}}^{\frac{\beta}{2}}d\tau_{1}\int_{-\frac{\beta}{2}}^{\frac{\beta}{2}}d\tau_{2}\,e^{i\omega\tau}\nonumber \\
&\!\left\langle\mathcal{T}_{\tau}\mathcal{J}_{i}\left(x,\tau\right)\mathcal{J}_{j}\left(x',0\right)
\mathbfcal{J}_{-1}\cdot{\bm\phi}\left(0,\tau_{1}\right)\mathbfcal{J}_{-1}\cdot{\bm\phi}\left(0,\tau_{2}\right)\right\rangle,
\end{align}
where $\beta=1/T$. As the scaling dimensions of $\mathbfcal{J}_{-1}\cdot{\bm\phi}$ and $\mathcal{J}_{i}$
are $1+\Delta$ and $1$, respectively, the four-point correlator in Eq.~(\ref{GT}) goes as $T^{2\Delta+4}$.  
Furthermore, because each space or time integral is directly proportional to $T^{-1}$, 
the four integrals $\int d\tau\int dx'\int d\tau_{1}\int d\tau_{2}$ in Eq.~(\ref{GT}) scale as $T^{-4}$.
Therefore, we conclude that for the quantum-Hall charge Kondo case 
\begin{eqnarray}
G_{ij}(T)-G_{ij}(0)\sim T^{2\Delta}\sim T^{\frac{4}{M+2}}\label{GTG0}
\end{eqnarray}
rather than $\sim T^{\Delta}$, as summarized in Table~\ref{table2}.

Such characteristic finite-temperature scaling behaviors of the conductances
are another hallmarks of the NFL's in the quantum-Hall charge Kondo criticality,
one of the key results of this study.
Particularly, for the $M=N=2$ case, our conclusion $G_{ij}(T)-G_{ij}(0)\sim T$ is in excellent agreement with
the recent experimental observation~\cite{Pierre} and numerical RG simulation~\cite{Sela}.

\indent\textcolor{blue}{\em Discussions.}---As we have established and compared in Tables~\ref{table1}~and~\ref{table2},
the zero-temperature conductances and their finite-temperature scaling laws sharply distinguish 
the quantum-Hall charge Kondo criticality from the quantum-dot spin Kondo criticality,
although they host the same family of NFL's.
Their distinct characteristics can be understood physically.
In a quantum-Hall charge Kondo transport~\cite{Pierre}, 
each lead acts as an independent channel of conduction electrons.
The conductance across the metallic island involves {\em two successive} scatterings with the NFL:
an electron tunneling in and out through different QPC's flips the charge pseudospin and then flips it back.
By contrast, in a quantum-dot spin Kondo transport~\cite{DGG1,DGG2,Bao}, 
the source and drain leads combine to form an effective channel.
The transmittance across the quantum dot involves {\em only one} scattering with the NFL:
an electron tunneling in and out from different leads exchanges spin with the dot once. 

Therefore, the conductance is related to the correlator
$\langle\mathcal{T}_{\tau}\psi_d(z)\psi^{\dagger}_s(z')\rangle$~\cite{Glazman} in the spin Kondo effect 
while to the correlator $\langle\mathcal{T}_{\tau}J_{i}(z)J_{j}(z')\rangle$ in the charge Kondo effect. 
Intriguingly, the two conductances can be unified as
\begin{align}\label{GSSSS}
{G}=\frac{e^{2}}{2h}\left[1-({S_{1/2}^{\lambda}/S_{0}^{\lambda}})/({S_{1/2}^{0}/S_{0}^{0}})\right],
\end{align}
where $S_{\mu}^{\nu}$ has been given above Eq.~(\ref{G1}), 
and $\lambda$ is determined by the scaling dimension of primary operator $\psi$ or $\mathcal{J}$. 
It follows that Eq.~(\ref{GSSSS}) reduces to Eq.~(\ref{G1}) for the charge Kondo case in which $\lambda=1$.
For the spin Kondo case, Eq.~(\ref{G1}) with $\lambda=1/2$ is exactly the conductance per spin in each channel~\cite{Borda,DGG1,DGG2,Bao}
(if the source-drain asymmetry prefactor~\cite{DGG2} is ignored).
Similar to the derivation of Eq.~(\ref{GTG0}), 
we can further understand the finite-temperature conductance for the spin Kondo case
using the leading irrelevant operator $\mathbfcal{J}_{-1}\cdot{\bm\phi}$. 
The temperature dependence arises from the first order correction 
$\sim\int d\tau\int d\tau'e^{i\omega\tau}\langle\mathcal{T}_{\tau}\psi(\tau)\psi^{\dagger}(0)\mathbfcal{J}_{-1}\cdot{\bm\phi}(\tau')\rangle$.
This yields $G(T)-G(0)\propto T^{\Delta}$~\cite{Affleck1},
as the scaling dimensions of $\psi$ and $\mathbfcal{J}_{-1}\cdot{\bm\phi}$ 
are $1/2$ and $1+\Delta$, respectively, and each time integral is directly proportional to $T^{-1}$.

As an intriguingly fact, the $M$-channel Kondo ground state amounts to
a decoupled ${Z}_M$ parafermion with a central charge $\frac{2M-2}{M+2}$. 
This generalizes the Majorana in the 2CK effect~\cite{Kivelson}
and explains the NFL behaviors at multi-channel Kondo fixed points~\cite{Affleck7}.
The Kondo problem involves $2M$ Dirac fermions, 
yet only the spin currents interact with the boundary spin, as seen in Eq.~(\ref{Ji}). 
This yields a RG flow merely in the $SU(2)_{M}$ sector from the weak-coupling to the Kondo fixed points;
the ground state degeneracy flows from $2$ to $2\cos\frac{\pi}{M+2}$, i.e., $S_{1/2}^{0}/S_{0}^{0}$.
The Kondo fixed points, independent of the exchange anisotropy, can be reached from an ingeniously designed limit,
e.g., the Toulouse-Emery-Kivelson limit~\cite{Kivelson} in which an $U(1)$ decouples.
As such, the RG flow takes place purely in the ${SU}(2)_{M}/U(1)$ sector, i.e., the ${Z}_M$ parafermionic theory. 
An ambitious goal in the future is to make such static parafermions mobile and manifest their non-abelian braiding statistics. 

{\em Note added.}---A complementary and independent experiment on the charge 3CK effect appeared~\cite{new} 
during the finalization of our manuscript. Their reported crossovers from the 3CK NFL to 2CK NFL and to 1CK FL, 
universal conductances across the 3CK and 2CK NFL's, 
and temperature scaling of the 2CK NFL conductance are consistent with our predictions.

\bibliographystyle{apsrev4-1}

\end{document}